\documentclass[aps,prl,twocolumn,groupedaddress]{revtex4}

\usepackage[utf8]{inputenc}
\usepackage[]{amsmath}
\usepackage[]{graphicx}
\usepackage{amssymb}
\usepackage{float}
\newcommand{\res}{\mathsf{res}}
\newcommand{\de}{\mathrm{d}}
\newcommand{\De}{\nabla}

\begin{document}


\title{A Variational Approach to Parameter Estimation in Ordinary Differential Equations}


\author{Daniel Kaschek$^{1,}$, Jens Timmer$^{1,2,3}$\\
	$^1$ Institute of Physics, Freiburg University, Freiburg, Germany\\
	$^2$ Freiburg Center for Systems Biology (ZBSA), Freiburg University, Freiburg, Germany\\
	$^3$ Freiburg Institute for Advanced Studies (FRIAS), Freiburg University, Freiburg, Germany}

\date{\today}


\begin{abstract}

	Ordinary differential equations are widely-used in the field of systems biology and chemical engineering to model chemical reaction networks. Numerous techniques have been developed to estimate parameters like rate constants, initial conditions or steady state concentrations from time-resolved data. In contrast to this countable set of parameters, the estimation of entire courses of network components corresponds to an innumerable set of parameters. The approach presented in this work is able to deal with course estimation for extrinsic system inputs or intrinsic reactants, both not being constrained by the reaction network itself. Our method is based on variational calculus which is carried out analytically to derive an augmented system of differential equations including the unconstrained components as ordinary state variables. Finally, conventional parameter estimation is applied to the augmented system resulting in a combined estimation of courses and parameters. The combined estimation approach takes the uncertainty in input courses correctly into account. This leads to precise parameter estimates and correct confidence intervals. In particular this implies that small motifs of large reaction networks can be analysed independently of the rest. By the use of variational methods, elements from control theory and statistics are combined allowing for future transfer of methods between the two fields.
\end{abstract}

\maketitle

\section{Introduction}
Frequently, signalling pathways and chemical reaction networks in systems biology are modelled by ordinary differential equations (ODE). In many cases, the reaction networks are open systems comprising external inputs like drug stimuli. The system is then modelled by a non-autonomous ODE.\\
Similarly, modules of reaction networks are open systems. The nodes they have in common with the surrounding network are not or not entirely determined by the module species. They can be considered as intrinsic inputs and again the system can be modelled by a non-autonomous ODE. An example for such a cross-talk can be found in \cite{kim_hidden_2007}. \\
While reaction rates and initial reactant concentrations form a countable set of parameters, inputs correspond to an innumerable set of parameters since in general, every function of time is possible as input and each function value at each time point is a free parameter. Commonly, if measurements for the inputs are available, non-parametric estimates like smoothing splines are employed to describe the input data \cite{raue_structural_2009}\cite{swameye_identification_2003}. Given the input, an objective function depending on rate parameters and initial values is defined and its minimum is approached by numerical optimization methods. In this way, the problem of infinitely many parameters is avoided. As we will show, one problem associated to this approach is that it does not account for the uncertainty present in the input. As a consequence, estimated parameter confidence intervals do not cover the actual variability, i.e.~they are too small.\\
Therefore, it is preferable to parametrize the input which is possible if certain knowledge about origin and processes underlying inputs is available. This enables a reasonable choice of basis functions and the parametrization becomes finite. Following this approach, the problem of erroneous confidence intervals is circumvented presuming that the input model is correct. However, this assumption is problematic if only sparse information about the inputs and few measurement points are available.\\
We propose to approach the problem of input parametrization by calculus of variations. In the Result section, the system's objective function used for ordinary parameter estimation is extended to a functional to be minimised. The original non-autonomous ODE is transformed into an augmented autonomous ODE. The result is interpreted and applied to simulated data. 

\section{Method}
\subsection{Derivation of the augmented ODE system}
In conventional parameter estimation, the objective function to be optimised is the likelihood function or the $\chi^2$ function. If a reaction network with species $y_{\mu}$, $\mu = 1, \dots, n$ and reaction parameters $p_k$, $k=1, \dots, r$, comprises inputs $x_{\nu}$, $\nu = 1, \dots, m$, the dynamics of the system is described by the model
\begin{equation}
	\dot y_{\mu}(t) = f_{\mu}\big(y(t), x(t), p\big),\quad y_{\mu}(0)=y_{\mu, 0},
	\label{eq:dynamics}
\end{equation}
with dynamic variables $y_{\mu}$ and time-dependent input functions $x_{\nu}(t)$, each of them collected in the vectors $y\in\mathbb R^n$ and $x\in\mathbb R^m$. In the following, the dependence on the whole course of $x$ will be emphasized by the notation $[x]$. Furthermore, it is assumed that the input function $x(t)$ is differentiable. Commonly used inputs like step functions or injections are rather distribution like than differentiable functions. However, it is assumed that on the physiological level the acting input is more accurately described by a differentiable function.\\
The $\chi^2$ objective function
\begin{align}
	\chi^2([x],p) = \sum\limits_{i, \mu} & \left(\frac{y_{\mu}(t_i, [x], p, y_0) -  y_{\mu, i}^{D}}{\sigma_{y_{\mu, i}^D}}\right)^2 +\nonumber\\
	& + \sum\limits_{i, \nu} \left(\frac{x_{\nu}(t_i) - x_{\nu, i}^{D}}{\sigma_{x_{\nu, i}^D}}\right)^2
	\label{eq:chisqu}
\end{align}
penalizes distances between species measurements $y_{\mu, i}^D$ and model prediction $y_{\mu}(t_i,[x],p,y_0)$ at time points $t_i$ quadratically and weighted by the measurement uncertainties $\sigma_{y_{\mu,i}^D}$. In addition, input measurements $x_{\nu, i}^D$ are compared with the input function values $x_{\nu}(t_i)$. In particular, $\chi^2$ is already a functional of $[x]$. In case of Gaussian noise, eq.~(\ref{eq:chisqu}) coincides with the maximum likelihood estimator.\\
Our aim is to find a unique input function which minimises the functional defined in eq.~(\ref{eq:chisqu}). To this purpose, we compute the first variation and check under which condition the first variation vanishes. See \cite{giaquinta_calculus_1996} for a general introduction to variational calculus as well as sections 1-2 in the supplement . For the first variation we obtain
\begin{align}
	\delta\chi^2 h = 2\sum\limits_{i} \bigg( & \res_{y}(t_i)\cdot\underbrace{\Phi(t_i)\int\limits_0^{t_i}\Phi^{-1}\De_x f h\,\de\tau}_{=\delta y h} + \nonumber\\
		& + \res_{x}(t_i)\cdot h\bigg).
	\label{}
\end{align}
The trajectory variation $\delta y h$ is derived by eq.~(\ref{eq:dynamics}) and is expressed by variation of constants: $\Phi(t)$ denotes the fundamental system of the homogeneous linear problem $\dot\phi = \De_y f \phi$ with the matrix $\De_y f$ of first derivatives of $f$ with respect to $y$ and $\De_x f h$ constitutes the inhomogeneity. Furthermore, a weighted residual function is defined as $\res_{y_{\mu}}(t_i) = \frac{y_{\mu, i}^D- y_{\mu}(t_i)}{\sigma_{y_{\mu, i}^D}^2}$, analogously $\res_{x_{\nu}}$. For a detailed derivation see sections 2-5 in the supplement.\\
Next, $h$ needs to be separated. Similarly to Euler-Lagrange's equation \cite{giaquinta_calculus_1996}, partial integration needs to be performed to extract $h$ from the integral. However, therefore the sum in eq.~(\ref{eq:chisqu}) needs to be extended to an integral, all time-discrete measurement points $y_{\mu, i}^D$ and $x_{ \nu,i}^D$ have to be replaced by continuous and differentiable data representations by means of a mapping $S: \mathbb R^N \rightarrow C^1(\mathbb R)$ from $N$ discrete values to a differentiable function. The resulting representations $S_{y_{\mu}}(t)$, $S_{x_{\nu}}(t)$ as well as $S_{\sigma_{y_{\mu}}}(t)$, $S_{\sigma_{x_{\nu}}}(t)$ need to be defined at least on a finite interval $[0, T]$ where $T$ denotes the latest time point to be considered. After partial integration, the first variation for the just defined time-continuous $\chi^2$ functional reads
\begin{align}
	\delta\chi^2 h = 2\int_0^T & \Bigg( \De_x f^* \underbrace{(\Phi^{-1})^*\int_t^T \Phi^*\res_y\,\de\tau}_{=:u} + \res_x \Bigg)\cdot h\,\de t,
	\label{eq:firstvariation}
\end{align}
with the auxiliary function $u$. The transpose is denoted by $^*$. The integral, i.e.~the first variation, vanishes for all choices of $h$ if and only if the integrand is zero, leading to eq.~(\ref{eq:xu}). The auxiliary function $u$ is equivalently expressed by its corresponding differential equation, eq.~(\ref{eq:u}). Here, it is used that $\Phi^{-1}$ is a fundamental system for $\dot\phi = -\nabla_y f \phi$ which follows from $\Phi$ being a fundamental system for $\dot\phi = +\nabla_y f \phi$. Together with eq.~(\ref{eq:dynamics}) we obtain:
\begin{align}
	0 &= \De_x f^* u + \res_x \label{eq:xu}\\
	\dot u &= -\De_y f^* u - \res_y \label{eq:u}\\
	\dot y &= f(y, x, p). \label{eq:y}
\end{align}
The right-hand sides of eqs.~(\ref{eq:u}-\ref{eq:y}) depend on state variables $y$, $u$, and $x$, the latter being constrained by eq.~(\ref{eq:xu}). Particularly, if the input enters linearly in the dynamics of the reaction network, $\De_x f$ is independent of $x$ and eq.~(\ref{eq:xu}) can be directly solved for $x$, i.e.~$x = S_x - S_{\sigma_x^2} \De_x f^* u$. However, even in the non-linear case, the implicit function theorem provides the possibility to check locally whether eq.~(\ref{eq:xu}) uniquely defines $x(u, y)$. For the discussion of a global version of this statement, see section 6 of the supplement.\\
From the definition of $u$ it follows that $u(T)=0$ needs to vanish at the final time point $T$. Hence, the augmented ODE system needs to satisfy two-way boundary conditions $y(0) = y_0$ and $u(T) = 0$. This fact constitutes a remarkable difference to the original initial value problem.\\

\subsection{Interpretation}
Starting from a dynamic system with inputs and measurements for both, state variables and inputs, we have derived differential equations for both of them. The original initial value problem has been transformed into a boundary value problem which is to be solved numerically. The solution trajectories $Y(t|p, y_0)=(y(t|p, y_0), x(t| p, y_0))$ minimise the $\chi^2$ functional for given dynamic parameters $p$ and initial values $y_0$. However, there is still notable freedom in the choice of data and uncertainty representations, denoted by $S_y$, $S_x$ and $S_{\sigma}$, which decides about the meaning of the solution trajectories.\\
One possibility to define time-continuous data representations $S_y$ and $S_x$ is smoothing splines. They constitute prior knowledge for each component about shape and time-scale of changes based solely on the measurement points. Also $S_{\sigma}$ needs to be chosen appropriately. Differences between model prediction and data prior are usually weighted by $w(t) = \frac{1}{S_{\sigma^2}(t)}$ at each time point $t$. Especially if data sampling is sparse, the data prior has larger uncertainty when far away from measurement time points. In this case, a reasonable choice of $w(t)$ is given by
\begin{equation}
	w_{\tau}(t) = \sum_i \frac{1}{\sqrt{2\pi\tau^2}}e^{-\frac{(t-t_i)^2}{2\tau^2}}\frac{1}{\sigma_i^2},
	\label{}
\end{equation}
i.e.~a sum of Gaussians located around the measurement points. The parameter $\tau$ is a measure for the correlation length of the data prior.\\
Once data and uncertainty representations are chosen, the solution trajectories $Y$ can be employed for conventional parameter estimation minimising
\begin{equation}
	\chi^2(p, y_0) = \sum_{\mu, i} \left(\frac{Y_{\mu, i} - Y_{\mu}(t_i| p, y_0)}{\sigma_{Y_{\mu, i}}}\right)^2
	\label{eq:augmchisqu}
\end{equation}
over the finite dimensional parameter space of $p$ and $y_0$. Note that the time-discrete $\chi^2$ function and the time-continuous $\chi^2$ functional do not coincide exactly. Thereby, different measures of optimality are applied to input functions and parameters. This difference is resolved in the asymptotic case of infinitely many measurement points.\\

The distinction between parameter estimation and input reconstruction has further implications on the estimation of uncertainty bounds. Confidence intervals can only be assigned to the dynamic parameters and initial conditions. In contrast, the input becomes a usual state variable by construction. For state variables, the confidence region in parameter space needs to be mapped to state space by prediction, i.e. by evaluating the model for different parameter values within the confidence region. This can e.g.~be realized by parameter sampling using MCMC methods. Alternatively, profile likelihoods can be employed \cite{Kreutz2011}.

\subsection{Technical remarks}

It is important for the interpretation of $x(t)$ as a species concentration that $x(t)>0$ for all times $t\in[0, T]$. This is not imposed a priori on the solution $x(t)$. Rather, it needs to be enforced by construction, analogously to the state variables in the ODE of the dynamic system. This can be realized by the following extension of the dynamic system,
\begin{align}
	\dot y &= f(y,x,p)\\
	\dot x &= -D(t) x,\label{eq:extension}
\end{align}
with a diagonal matrix $D(t) = \textsf{diag}\big(d_1(t), \dots, d_{m}(t)\big)$ of new inputs $d_1, \dots, d_m$. By construction, $x$ can not change sign over time. While the prior course for $D(t)$ is rather clear, i.e. $d_{\nu}(t) = -\frac{\dot S_{x_{\nu}}}{S_{x_{\nu}}}$, the uncertainty prior constitutes a subjective choice. It reflects the belief how fast $x(t)$ can change at most. Besides enforcing positivity of the input, the extension by eq.~(\ref{eq:extension}) presents a workaround for dealing with non-linear inputs because the new input variables $d_{\nu}$ enter linearly and the old inputs $x_{\nu}$ become regular state variables.\\

If $f$ depends linearly on $x$, eq.~(\ref{eq:xu}) can be solved for $x$ explicitly. This ensures computational efficiency. In the non-linear case, matrix inversion has to be performed in each evaluation step of the ODE which might slow down the computation of the solution remarkably. Alternatively to the introduction of new input variables, eq.~(\ref{eq:extension}), the computationally intensive approach can be avoided by a change of variables. This is possible if state variables and input variables factorize, i.e. if
\begin{align}
	f_{\mu}(y, x, p) = \sum_{\nu = 1}^m g_{\mu\nu}(y, p) \tilde x_{\nu} + g_{\mu,0}(y,p),\quad\mu = 1, \dots, n,
	\label{}
\end{align}
where $g_{\mu\nu}$ and $g_{\mu,0}$ do not depend on the input variables which have been transformed to $\tilde x = \varphi(x, p)$ by a coordinate transformation $\varphi$. Examples could be $\varphi(x) = x^2$ or $\varphi(x, K_D) = \frac{x}{K_D + x}$ for a bimolecular or an enzymatic reaction, respectively. The possibility of a change of variables covers a broad range of biologically relevant reaction networks.\\

Although computation for linear input is remarkably faster than for non-linear input, it is still slower than solving an initial value problem. On the other hand, the solution of the boundary value problem is already optimal with regard to the input course. Therefore, computing time has to be compared to the time a parameter optimization algorithm takes to estimate the parametrized input course. The comparison will strongly depend on the number of parameters that are necessary to parametrize the partially unknown input. So far, there has not been a comprehensive study comparing the two methods.

\section{Application to simulated data}
The approach is applied to the following toy model:
\begin{align}
	\begin{array}{ccc}
		& x & \\
		& \downarrow & \\
		A & \rightleftharpoons & B.
	\end{array}
	\label{}
\end{align}
The forward reaction $A\rightarrow B$ is mediated by $x$ while the back reaction $B\rightarrow A$ is unaffected by the input $x$. According to eqs.~(\ref{eq:xu}-\ref{eq:y}), the augmented ODE system for $A$, $B$ and $x$ is given by
\begin{align}
	\dot A &= -k_1 A x + k_2 B\\
	\dot B &= k_1 A x - k_2 B\\
	\dot u_A &= k_1 x(u_A-u_B) - \frac{A - S_A}{S_{\sigma_A^2}}\\
	\dot u_B &= -k_2 (u_A - u_B) - \frac{B - S_B}{S_{\sigma_B^2}}
	\label{}
\end{align}
with the auxiliary state variables $u_A$, $u_B$, the data representations $S_A$, $S_B$ and the uncertainty representations $S_{\sigma_A^2}$ and $S_{\sigma_B^2}$. The input $x$ is related to the other state variables by $x = S_x + S_{\sigma_x^2} k_1 A(u_A -u_B)$. Several input functions $x$ have been chosen for data generation, among them an exponential decay, $x\sim e^{-\alpha t}$, an activation dynamics with a slow decay after a fast increase, $x\sim e^{-\alpha t} - e^{-\beta t}$ with $\alpha>\beta$, and a Gaussian input, $x\sim\frac{1}{\sqrt{2\pi\tau^2}}e^{-\frac{(t-t_{\textrm{dip}})^2}{2\tau^2}}$. The example is numerically implemented in C and in R \cite{R2011}. Optimization is performed by a Gauss-Newton algorithm for nonlinear least-squares estimation.\\

The purpose of this section is to compare parameter estimation for the variational and the fixed input approach. The input data prior, i.e.~the smoothing spline through the simulated input data points, is employed as input function for the fixed input approach.\\

\begin{figure}[h]
	\begin{center}
		\includegraphics[width=0.48\textwidth]{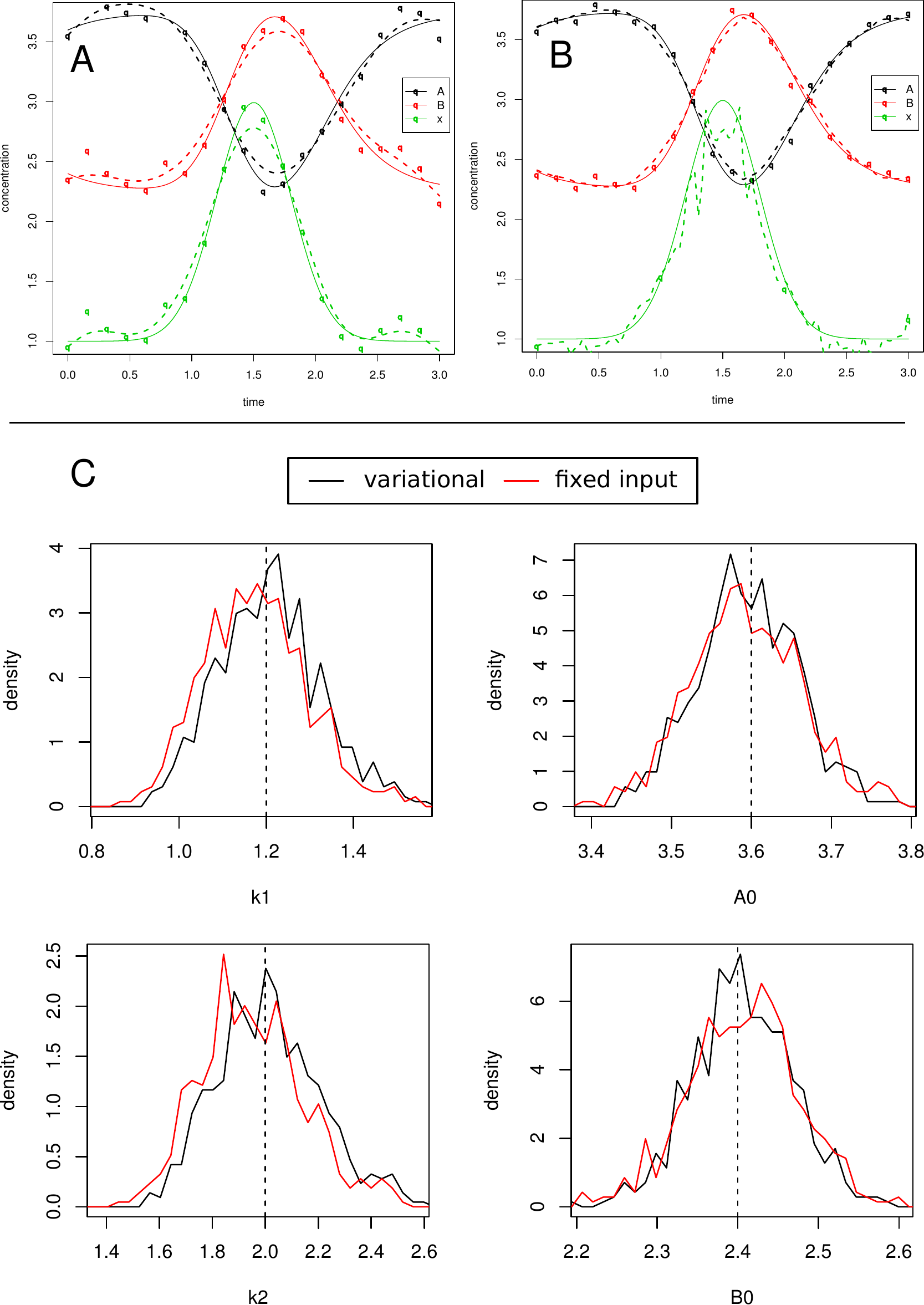}
	\end{center}

	\caption{
	(A-B) Simulated data for the species $A$, $B$ and the input $x$. True time courses are denoted by continuous lines. Data points are subject to Gaussian noise with $\sigma=0.1$. (A) Data representations are indicated as dashed lines, (B) solution trajectories after parameter estimation are shown as dashed lines. (C) Comparison of parameter distributions obtained from 1000 repetitions of data generation and parameter estimation for the variational and the fixed input approach.
	}
	\label{fig:splineAndFit20}
\end{figure}

\begin{figure}[h]
	\begin{center}
		\includegraphics[width=0.48\textwidth]{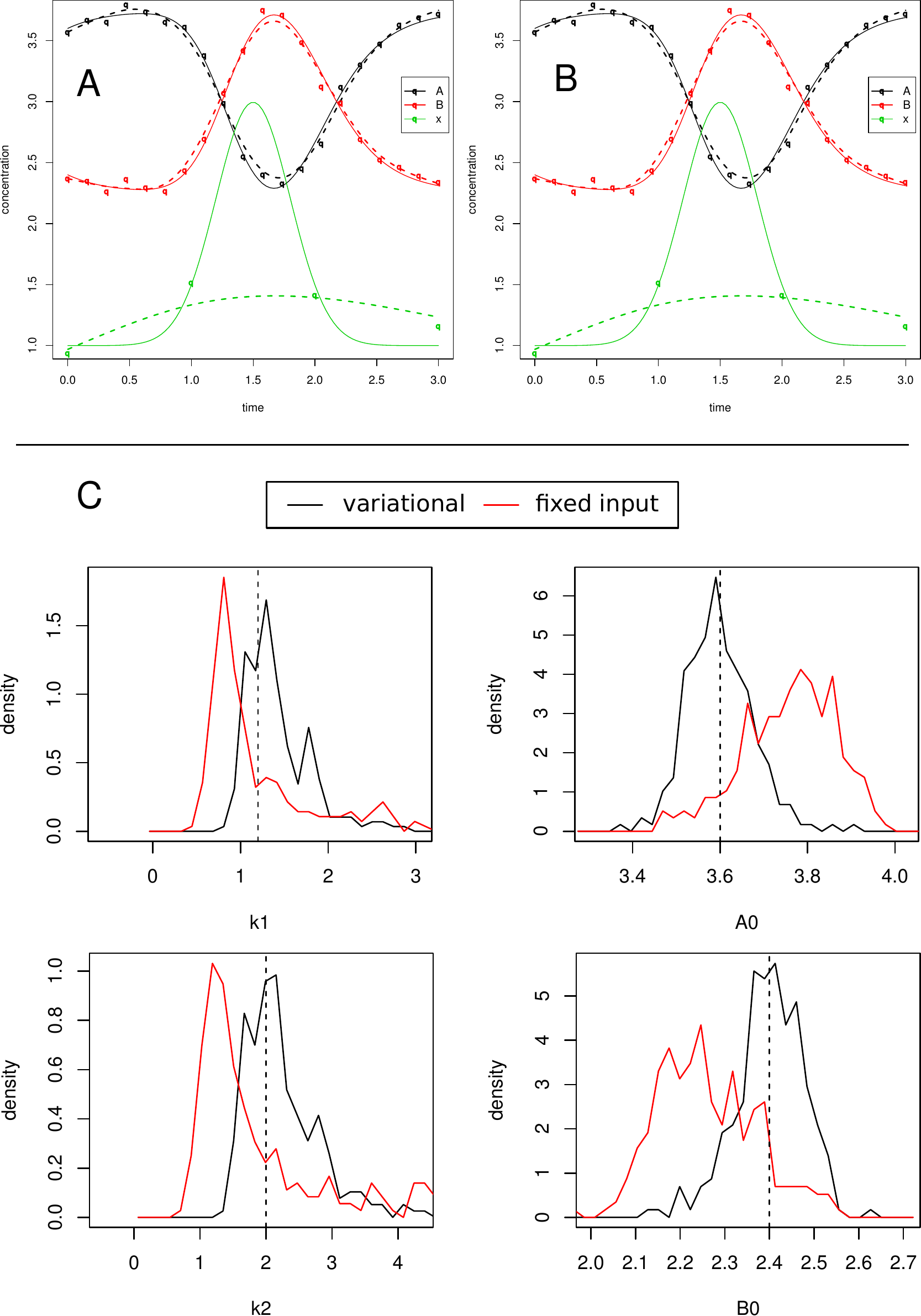}
	\end{center}
	\caption{(A-B) Input reconstruction -- simulated data for the species $A$, $B$ and the input $x$. True time courses are denoted by continuous lines. Data points are subject to Gaussian noise with $\sigma=0.1$. For the input, only 4 data points are provided. (A) Data representations are indicated as dashed lines, (B) solution trajectories after parameter estimation are shown as dashed lines. (C) Comparison of parameter distributions obtained from 1000 repetitions of data generation and parameter estimation for the variational and the fixed input approach.}
	\label{fig:splineAndFit4}
\end{figure}

Examples with Gaussian input are depicted in Figures~\ref{fig:splineAndFit20} and \ref{fig:splineAndFit4}. All components, $A$, $B$, and $x$ depicted in Figure~\ref{fig:splineAndFit20}A-B have been measured at 20 time points. In this case of dense sampling, the data priors, charted as dotted lines in Figure~\ref{fig:splineAndFit20}A, come close to the estimated time-courses, charted as dotted lines in Figure~\ref{fig:splineAndFit20}B. This is reflected in the distributions of the parameter estimates in Figure~\ref{fig:splineAndFit20}C: for the same set-up, 1000 noise realizations have been generated and the variational approach has been used for parameter estimation. In order to compare the result with the fixed input approach, the data prior of $x$ has been employed as input and conventional parameter estimation has been performed. Hence, in the setting of dense sampling and small noise, both estimation approaches perform equally in terms of accuracy and precision.\\

A rather different situation is depicted in Figure~\ref{fig:splineAndFit4}A-B. The input $x$ is measured at four time points only, leading to a poor data prior shown as green dotted line in Figure~\ref{fig:splineAndFit4}A. Like before, the species $A$ and $B$ have been measured at 20 time points. Most of the information about the dynamics of the input is encoded in these measurements. The correlation time $\tau$ has been chosen to be much smaller than the distance between time points allowing for much interstitial variability. The resulting trajectories $Y$ after parameter estimation are shown as dotted lines in Figure~\ref{fig:splineAndFit4}B. The true input curve is reconstructed almost entirely. The noticeable fluctuations are caused by coincidental noise correlations between species $A$ and $B$: simultaneous deviations from the true course in opposite directions lead to immediate breakouts of the reconstructed input.\\
Also for this set-up, 1000 noise realizations have been generated for the comparison of the variational and the fixed input approach. The parameter and initial value distributions for both approaches are shown in Figure~\ref{fig:splineAndFit4}C. Since the true input can be reconstructed, the variational approach is able to estimate all parameters accurately. In contrast, when the input is fixed to the apparent input data prior, parameter estimation leads to biased parameter estimates.\\ 

Finally, we investigated the coverage probability \cite{suess_introduction_2010} of the confidence intervals derived from the variational and the fixed input approach: for each simulated data set, parameter estimation is performed, confidence intervals are computed and the information if the true parameter value is situated within the 68\%/90\% confidence interval is collected. This information is cumulated over many runs of data generation.\\
\begin{figure*}[t]
	\begin{center}
		\includegraphics[width=\textwidth]{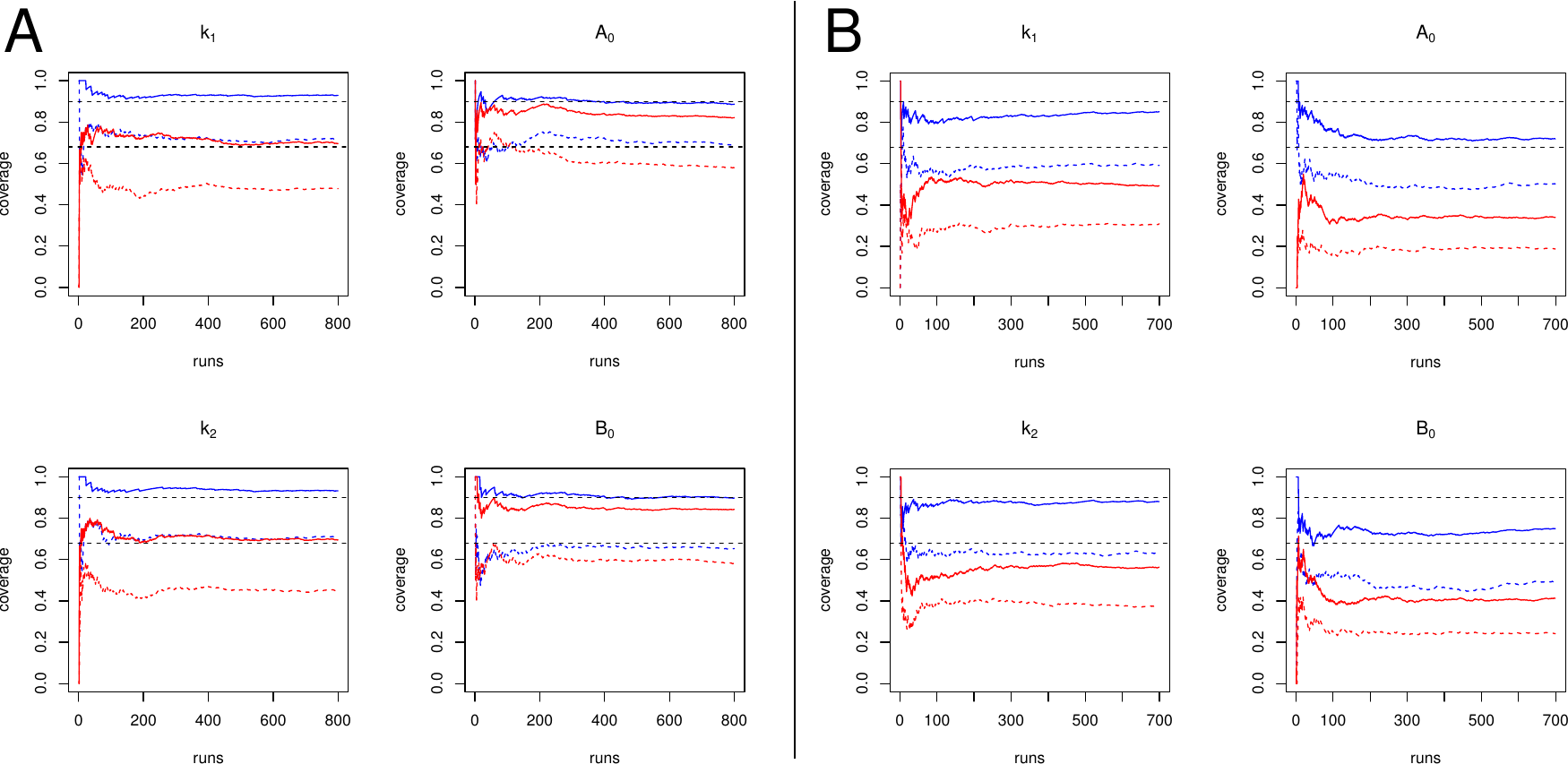}
	\end{center}
	\caption{Coverage for variational approach (blue) and fixed input approach (red). Continuous lines correspond to a 90\% coverage probability, dashed lines to 68\%. Both probabilities are indicated as black dashed horizontal lines. (A) Example with 20 input measurement points, (B) example with 4 input measurement points.}
	\label{fig:coverage}
\end{figure*}
Figures~\ref{fig:coverage}A and \ref{fig:coverage}B show the results for Gaussian input with 20 input measurements and 4 input measurements respectively. In each case, 20 measurement points have been provided for each of the species $A$ and $B$.\\
For both estimation approaches, confidence intervals of estimated parameters and initial values have been produced by means of the profile likelihood approach \cite{murphy_profile_2000} with respect to eq.~(\ref{eq:augmchisqu}).\\
For the set-up with 20 input measurement points, both estimation approaches provide accurate estimators with similar variances as confirmed by Figure~\ref{fig:splineAndFit20}. However, as Figure~\ref{fig:coverage} shows, the coverage differs significantly between the two approaches. Confidence intervals for $k_1$ and $k_2$ are systematically underestimated by the fixed input approach. The variational approach in contrast is able to correctly take the degrees of freedom in the input into account. Thus, the coverage is close to the expected values.\\
For the set-up with 4 input measurement points, the variational approach performs significantly better than the fixed input approach with respect to coverage. However, also the variational approach produces confidence intervals that are slightly too small for the dynamic parameters $k_1$ and $k_2$, Figure~\ref{fig:coverage}B left, and too small for the estimated initial values, Figure~\ref{fig:coverage}B right. The reason for this behaviour is a combination of the small correlation length $\tau$ and the objective function given by eq.~(\ref{eq:augmchisqu}). Short values of $\tau$ allow that the input function has fast fluctuations. Especially around the input measurement points, function values tend to punctually approach the measured values, favoured by the time-discrete objective function. Since these fluctuations occur at a short time scale, it has little influence on the course of $A$ and $B$ and thereby, estimation of the dynamic parameters is almost unaffected.\\
This case shows that $\tau$ needs to be chosen appropriately for the problem: small for comprehensive input reconstruction and larger for propagation of uncertainties. A second possibility would be to adapt statistical results for conventional parameter estimation to the case of time-continuous objective functions.

\section{Conclusion}\label{sec:conclusion}
In many applications, it is difficult to guess a proper input model because input data is not available or too noisy. Instead of parametrizing the input, we employed variational calculus to transform the ODE into an augmented system of ODEs describing the original and the input components. The solution of this system minimises the $\chi^2$ functional which plays a central role and is directly associated to the objective function of the original estimation problem. Since the extension of the $\chi^2$ function to the $\chi^2$ functional is not unique, the new functions, i.e.~continuous data and uncertainty representations, need to be chosen intentionally. To this end we propose smoothing splines that have a concrete interpretation as data priors. Especially in the case of sparse sampling we propose to use weighting functions for the uncertainties. By this means, existing measurement points are taken into account and the course between time points is not excessively constrained by the data prior.\\

In the field of control theory and optimal control, so called cost functionals take the role of our $\chi^2$ functional. Once chosen the appropriate $\chi^2$ functional, our approach to input estimation can be embedded in the general framework of Pontryagin's minimum principle \cite{kirk_optimal_2004} and eqs.~(\ref{eq:u}-\ref{eq:y}) can be identified as a Hamiltonian system. \\

We showed that our combined variational approach to parameter estimation enables the assembly of all information present in species and input measurements. By this means, it accounts properly for variability in the input due to measurement uncertainties and produces correct confidence bounds. Depending on the situation, the combination of all information leads to comprehensive reconstruction of the input curves. Information about the dynamics of the input can be concentrated in the species measurements like Figure~\ref{fig:splineAndFit20} shows. In such cases our approach clearly outperforms conventional approaches. The variational method is even applicable if no input measurements are available or if species are partially unobserved. A prominent example where the presented method could be applied is the PI3K/AKT/mTOR pathway \cite{Pezze2012}. Even though various mTOR complexes and their phosphorylated states can be measured, it is not clear how they mediate AKT activation. By applying the variational method to AKT data, it would be possible to reconstruct the required mediator and subsequently relate it to mTOR complex measurements.\\

A completely different field of application is network modularization. The entire network can be dissected preferably at nodes where measurements are available. These nodes are then treated as independent inputs thus disentangling the network. In this way, the number of equations the variational approach has to deal with is kept small and computational efficiency is ensured.\\ 

A further step after the introduction of a time-continuous objective function would be to use the same function for parameter estimation. The time-continuous version of the objective function is closely related to the original function. Therefore, we are confident that it is possible to endow the time-continuous objective function with statistical meaning. This would not only allow for employing the same objective for parameter estimation and input reconstruction in our application. It would also enable the transfer of many more results from control theory and make it suitable for statistical inference.

\section*{Acknowledgements}
	The authors thank Raphael Engesser, Clemens Kreutz and Jan Hasenauer for their advice and valuable discussions. This work has been supported by the German Federal Ministry for Education and Research programme Medical Systems Biology SARA (0315394E).

\bibliography{Publication-Variational}

\end{document}